\documentclass[runningheads,a4paper]{llncs}

\usepackage{multirow}
\usepackage{amssymb}
\usepackage{graphicx}
\usepackage{tabularx,booktabs,makecell}
\usepackage{tikz}
\usepackage{amsmath}
\usepackage{graphicx}
\usepackage{subcaption}
\newcolumntype{L}[1]{>{\raggedright\arraybackslash}p{#1}}

\usepackage{tabularx}
\usepackage[table]{xcolor} 
\usepackage{doi}
\usepackage{array}          

\usepackage{url}
\urldef{\mailsa}\path|{alfred.hofmann, ursula.barth, ingrid.haas, frank.holzwarth,|%
\urldef{\mailsb}\path|anna.kramer, leonie.kunz, christine.reiss, nicole.sator,|%
\urldef{\mailsc}\path|erika.siebert-cole, peter.strasser, lncs}@springer.com|%
\newcommand{\keywords}[1]{\par\addvspace\baselineskip
\noindent\keywordname\enspace\ignorespaces#1}
\newcommand{\samethanks}{\footnotemark[\value{footnote}]}
\usepackage[markup=underlined]{changes}

\begin{document}

\mainmatter  

\title{Large Language Model based Smart Contract Auditing with LLMBugScanner}


%
%
\author{Yining Yuan\thanks{These authors contributed equally to this work.} \and
        Yifei Wang\samethanks \and
        Yichang Xu \and
        \\
        Zachary Yahn \and
        Sihao Hu \and
        Ling Liu \thanks{Corresponding author: Ling Liu}}

\institute{School of Computer Science, 
           Georgia Institute of Technology, \\
           Atlanta, GA 30332, 
           United States\\
           \email{\{yyuan394, ywang4343, xuyichang, zachary.yahn, sihaohu, ll72\}@gatech.edu}}

\authorrunning{Y. Yuan, Y. Wang et al.}


%
%

\toctitle{Lecture Notes in Computer Science}
\tocauthor{Authors' Instructions}
\maketitle

\begin{abstract}
This paper presents LLMBugScanner $-$ a large language model (LLM) powered approach to smart contract vulnerability detection with fine-tuning and ensemble learning. 
First, the paper identifies the challenges of leveraging for auditing and bug detection of smart contract programs. For example, different pre-trained LLMs may have different reasoning capabilities, and a single LLM may not be consistently effective for all vulnerability types and/or all type of smart contract programs. Such problems also persist across fine-tuned LLMs. 
To address such challenges, this article explores the adaptation of domain knowledge and ensemble learning and integrates these two synergistic strategies for effective and enhanced generalization performance of smart contract vulnerability detection. With domain knowledge adaptation, we fine-tune LLMs on complementary datasets to provide both general code interpretation and instructional supervision, using parameter-efficient techniques to reduce computational cost. With LLM-ensemble, we capitalize on the complimentary wisdom of diverse LLMs to generate improved reasoning results through consensus based conflict resolution.    
Extensive experiments are performed on multiple popular LLMs. We compare LLMBugScanner with both individual pre-trained LLMs and their fine-tuned versions in terms of their smart contract vulnerability detection performance.   
Our measurement results show that LLMBugScanner delivers consistent accuracy gains and better generalization, highlighting LLMBugScanner as a principled, cost-effective, and extensible framework for smart contract auditing. 

\end{abstract}

\section{Introduction}

Smart contracts are self-executing programs that are managed and executed on the blockchain, enabling decentralized applications such as DeFi and NFTs to operate without intermediaries, collectively managing digital assets~\cite{khan2021blockchain}. This immutability (the contract code cannot be directly modified once deployed) serves as a foundation for trust and transparency in decentralized systems. However, it also faces critical risk: Even small bugs in smart contracts can cause catastrophic consequences, including theft, token devaluation, or permanent/unauthorized locking of funds. Research analysis confirms that common programming pitfalls in Ethereum contracts create exploitable security risks that can enable attackers to steal assets or disrupt functionality~\cite{iuliano2024smartcontractvulnerabilities,vidal2024vulnerabilitydetection}. A quintessential example is the 2016 The DAO exploit, where a reentrancy bug allowed attackers to siphon off about \$60 million worth of Ether, resulting in a controversial hard fork of the Ethereum blockchain~\cite{Mehar2017DAO}. More recently, DeFi platforms have seen security breaches on an even larger scale, underscoring the immense financial risks involved~\cite{hu2023bert4eth,hu2024zipzap}. As a result, there's growing demand for reliable detection tools that can identify vulnerabilities before deployment. A systematic literature review identified nearly 192 different types of Ethereum smart contract vulnerabilities and over 200 detection tools~\cite{iuliano2024survey}.

Traditional static and dynamic program analyzers~\cite{mythril,feist2019slither,tsankov2018securify,manticore} form the core of current auditing pipelines. 
However, prior empirical studies 
reveal persistent weaknesses, including high rates of false-positives, incomplete coverage of complex control flows, function-specific blind sports, and limited robustness to new vulnerability patterns~\cite{iuliano2024smartcontractvulnerabilities,vidal2024vulnerabilitydetection}. These limitations arise because rule-based analyzers depend on pre-defined patterns and fail to capture the actual logic or intent expressed in the code~\cite{hu2023large}.

Recent advances in large language models (LLMs) present an opportunity for developing a new paradigm for vulnerability detection~\cite{AlfredPros_CodeLLaMa7BInstructSolidity_2024}. 
We argue that LLMs offer key advantages over static analyzers through their generality, contextual reasoning, and interpretability. 
For example, studies show that LLMs trained on code corpora achieve superior semantic code understanding and generation~\cite{roziere2023code}.
Unlike rule-based tools, LLMs holds the potential 
to detect both known and new vulnerabilities, by reasoning 
across different semantics of smart contract programs, from logic, comments, to specifications, reducing false negatives in complex issues, e.g., reentrancy or access control flaws. Moreover, LLMs can explain their reasoning in natural language, providing interpretable insights that help auditors and developers validate and prioritize bug fixes. 
Several projects~\cite{hu2023large,gptscan} have pioneered the use of LLMs as foundational components in smart contract auditing frameworks. 
Despite these advances, existing LLM-based approaches remain limited. First, single-model fine-tuning can overfit to specific vulnerability types or datasets, leading to inconsistent predictions across domains~\cite{du2024generalization}. 
Second, different LLMs exhibit different reasoning capabilities and no single LLM or its fine-tuned version can correctly and persistently outperform others for smart contract vulnerability detection~\cite{hu2023large}.  

To address these issues, we present LLMBugScanner, an LLM-based smart contract vulnerability detection framework with parameter-efficient adaptation. 
The design of LLMBugScanner explores the Synergism of domain knowledge adaptation and LLM-ensemble to improve its generalization performance. First, we incorporate \textit{Domain knowledge adaptation} with heterogeneous smart contract benchmark datasets and parameter-efficient optimization~\cite{hu2021lora}. This allows LLMBugScanner to capture diverse semantic patterns of smart contract programs, while keeping fine-tuning of LLMs resource-efficient. Second, we develop an \textit{LLM-Ensemble} approach to combine the reasoning results of multiple independently fine-tuned LLMs to fortify the robustness of the vulnerability detection of LLMBugScanner across diverse vulnerability categories. 
%
We performed experiments on 108 real-world smart contracts, all of which were reported to contain vulnerabilities in the Common Vulnerabilities and Exposures (CVEs) database~\cite{cve}. 
Empirically, LLMBugScanner achieves a 60\% top 5 detection accuracy on 108 real-world vulnerable contracts from the curated subset of CVE-labeled Solidity contracts\cite{hu2023large}, outperforming single-model baselines by 19\% while maintaining efficiency. The results highlight LLMBugScanner as a scalable and robust approach for smart contract vulnerability detection.

\begin{table}[htbp]
\centering
\scriptsize 
\renewcommand{\arraystretch}{1.0}
\setlength{\tabcolsep}{4pt}
\caption{
\textbf{Categorization of smart contract vulnerabilities.}
Adapted from Atzei et al.~\cite{atzei2017survey} and extended with insights from later works~\cite{luu2016making,zheng2020overview}.
Vulnerabilities are grouped by layer (Solidity language, the Ethereum Virtual Machine (EVM), and blockchain design) to provide a holistic view of common pitfalls and attack surfaces in Ethereum smart contracts.
}
\label{tab:full_vulnerabilities}

\begin{tabularx}{\textwidth}{p{1.5cm} p{3.8cm} X}
\toprule
\textbf{Layer} & \textbf{Vulnerability} & \textbf{Description} \\
\midrule

\multicolumn{3}{l}{\textbf{Solidity programming language}} \\[2pt]
\cmidrule(l){1-3}
 & Denial of Service (block gas limit) & Transactions fail if gas usage exceeds the block gas limit (e.g., large loops or arrays). \\
 & Denial of Service (failed call) & Repeated failing external calls in loops can block contract progress. \\
 & Randomness via \texttt{blockhash} & \texttt{blockhash} is predictable and miner-influenced; unsafe for randomness. \\
 & \texttt{tx.origin} misuse & Using \texttt{tx.origin} for authentication enables phishing-style attacks. \\
 & Integer over/underflow & Arithmetic beyond type bounds can corrupt contract state. \\
 & Re-entrancy & External calls before state updates enable repeated withdrawals (e.g., DAO attack). \\
 & Mishandled exceptions & Failing to propagate or handle call errors creates hidden faults. \\
 & Gasless \texttt{send} & Fallbacks exceeding the 2300 gas stipend fail, potentially exploitable. \\
 & Gas-costly patterns & Inefficient loops or storage use waste gas (e.g., flagged by GASPER). \\
 & Call to unknown & Fallback invoked on undefined functions or raw Ether transfers. \\
& Hash collision in \texttt{abi.encodePacked} &
Using \texttt{abi.encodePacked()} with variable-length arguments (e.g., strings) can produce identical byte sequences for different inputs (e.g., "abc","def" vs. "ab","cdef"), leading to hash or signature collisions in verification logic. \\

 & Insufficient gas griefing & Starving subcalls of gas forces upstream reverts. \\
 & Unprotected Ether withdrawal & Missing access control permits unauthorized withdrawals. \\
 & Floating pragma & Unpinned compiler version risks inconsistent builds or compiler bugs. \\
 & Default visibility & Omitted visibility exposes unintended public functions. \\
 & \texttt{DELEGATECALL} misuse & Executes foreign code in caller context, risking state contamination. \\
 & Unprotected \texttt{selfdestruct} & Allows deletion or asset loss without proper authorization. \\

\addlinespace[3pt]
\multicolumn{3}{l}{\textbf{Ethereum Virtual Machine (EVM)}} \\[2pt]
\cmidrule(l){1-3}
 & Immutable bugs & Deployed bytecode is immutable; defects are permanent. \\
 & Ether lost in transfer & Sending to non-code or orphan addresses irreversibly loses funds. \\

\addlinespace[3pt]
\multicolumn{3}{l}{\textbf{Blockchain design layer}} \\[2pt]
\cmidrule(l){1-3}
 & Timestamp dependency & Miners can bias timestamps to influence contract logic. \\
 & Transaction-ordering dependency & Ordering/front-running exploits race conditions in the mempool. \\
 & Lack of transactional privacy & Public on-chain data can be inferred and misused. \\
 & Untrusted oracles & Malicious or erroneous off-chain sources can corrupt on-chain state. \\
\bottomrule
\end{tabularx}
\end{table}


\section{Background and Related Work}
\subsection{Smart Contract Basics}

Smart contracts are self-executing programs stored on blockchain platforms (e.g., Ethereum), designed to enforce agreements without intermediaries~\cite{khan2021blockchain}. Once deployed, the contract code is immutable and publicly accessible, providing transparency but also meaning any vulnerabilities cannot be patched in place~\cite{nikolic2018finding}. This immutability, combined with the high financial stakes, makes smart contract flaws particularly critical~\cite{rameder2022review}. Unfortunately, real-world smart contracts have been riddled with bugs and security vulnerabilities. Researchers have documented a broad taxonomy of common smart contract vulnerabilities from reentrancy bugs to integer overflows, that attackers can exploit to steal funds or disrupt contract logic~\cite{luu2016making,atzei2017survey}. To contextualize these risks, Table~\ref{tab:full_vulnerabilities} categorizes common smart contract vulnerabilities across different layers of the Ethereum ecosystem, offering a structured overview that highlights the diverse sources of attack surfaces and motivates the need for automated vulnerability detection. 


Several tools and frameworks now exist to detect known vulnerability patterns in smart contracts. Early security analyzers like Oyente showed that automated symbolic analysis could find prevalent bugs such as reentrancy and timestamp dependence~\cite{luu2016making}. Subsequent static analyzers formalized vulnerability patterns as compliance and violation conditions to flag insecure code constructs~\cite{tsankov2018securify}. These analyzers successfully identified many flawed contracts in the wild. For instance, Nikolić et al.\ scanned roughly one million Ethereum contracts and uncovered tens of thousands of vulnerable instances of greedy or suicidal contracts that lock funds or destroy ownership improperly~\cite{nikolic2018finding}. The cumulative lesson from this line of work is that smart contracts are prone to recurring bug types despite their simplicity, and pattern-based detectors can catch certain classes of bugs~\cite{nikolic2018finding}. However, traditional approaches rely heavily on expert-crafted rules and formal properties. This reliance yields a fundamental limitation: when contracts deviate from known patterns or when new vulnerability variants emerge, purely rule-based tools struggle to keep up~\cite{rameder2022review}. In practice, popular analyzers like Mythril or Slither report high false positive rates and often miss nuanced flaws that do not match their predefined signatures~\cite{durieux2020empirical}. The need for more flexible, learning-based analysis techniques has become apparent as the complexity and volume of smart contracts grow~\cite{rameder2022review}.

\subsection{LLM-based Vulnerability Detection}

A key factor driving the paradigm shift from traditional approaches~\cite{li2018vuldeepercker,zou2020muvuldeepercker} in software vulnerability detection is the proven ability of code-oriented LLMs to effectively model both the semantic meaning and structural patterns of source code~\cite{roziere2023code}. However, early evaluations show that generic LLMs can sometimes flag vulnerability patterns, frequently produce false positives~\cite{david2023manualaudit}, resulting in a low precision and necessitate exhausting manual verification efforts~\cite{hu2023large}. To mitigate the above-mentioned issues, multi-agent frameworks that enable LLMs to assume different roles have been introduced~\cite{hu2024survey}. A representative example in smart contract vulnerability detection is GPTLens~\cite{hu2023large}, which decomposes the detection task into two stages with distinct agent responsibilities: auditors that propose potential vulnerabilities, and a critic that refines the outputs and helps reduce false positive rates~\cite{hu2023large}. Later, LLM-SmartAudit employs a multi-agent conversational structure: individual agents specialize (e.g. code reader, logic checker), cross-validate each other’s outputs, and jointly produce a final audit report. Experiments show significant improvements over traditional tools~\cite{llm-smartaudit}.  
Smartify extends this approach across multiple smart-contract languages (Solidity and Move) by deploying an ensemble of fine-tuned agents specialized in detection and repair tasks, outperforming several standalone LLMs~\cite{smartify}.  
Recent surveys of LLM-based multi-agent systems also highlight that such collaborative approaches can reduce hallucination, offer richer reasoning paths, and mimic human auditing workflows more closely~\cite{survey-llm-mas}.  
These multi-agent approaches open a new frontier: rather than relying on one monolithic LLM, we can compose multiple specialized agents or models that reason and critique each other, potentially improving robustness and interpretability in contract auditing.

\section{Motivation}

\subsection{Domain Knowledge Adaptation}

Domain-specific fine-tuning has become a central strategy in adapting pre-trained LLMs to specialized downstream tasks. It enables a model to transfer general code understanding ability and programming knowledge acquired during pre-training to specific domains~\cite{devlin2019bert,zhuang2020comprehensive}. In our case, fine-tuning serves to adapt a general model into a specialized auditor capable of identifying vulnerabilities for smart contracts.
\begin{figure}
    \centering
    \includegraphics[width=1\linewidth]{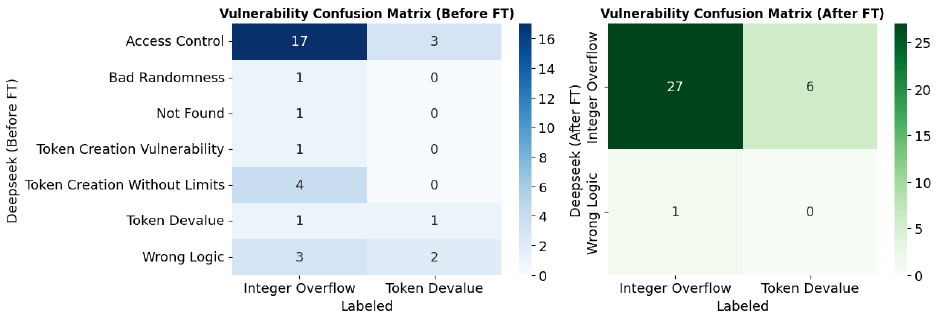}
    \caption{Confusion matrices of vulnerability classification before and after fine-tuning (FT). The left matrix shows DeepSeek’s baseline predictions, where diverse categories such as “Access Control,” “Token Creation Vulnerability,” and “Wrong Logic” were often misclassified as “Integer Overflow” or “Token Devalue.” After fine-tuning (right), the model demonstrates substantially improved discrimination, with most cases of “Integer Overflow” correctly identified and reduced confusion across categories.}
    \label{fig:confusion}
\end{figure}
Baseline LLMs used in a zero-shot configuration frequently misclassify safe constructs as unsafe or overlook non-obvious security flaws~\cite{hu2023large}. This stems from the model’s lack of exposure to Solidity’s nuanced semantics, where superficially similar constructs can differ drastically in safety. For example, Kharkar et al.~\cite{kharkar2022learning} report that an unadapted transformer mislabels many benign patterns as vulnerabilities due to spurious correlations, while a fine-tuned model more accurately separates true bugs from noise.

Figure~\ref{fig:confusion} contrasts the a baseline LLM and our fine-tuned LLM on labeled smart contract data. The empirical results show that fine-tuning substantially aligns model predictions with true vulnerability types, reducing confusion between unrelated categories. For instance, the baseline model frequently misinterprets integer overflow vulnerabilities as benign logical issues or misclassifies unchecked call patterns. After fine-tuning, the Auditor correctly identifies these vulnerabilities with significantly fewer false positives, reflecting improved semantic discrimination.


\subsection{Efficient Ensemble Learning}

Fine-tuning an LLM on vulnerability-detection data improves domain specificity, yet relying on a single adapted model can leave systematic biases toward certain vulnerability types or datasets. Empirical studies report large run-to-run variability and low inter-model agreement for deep learning–based vulnerability detectors, and show that models struggle on under-represented CWEs and out-of-distribution (OOD) code~\cite{steenhoek2023empirical,chen2023diversevul,du2024generalization}.

Ensemble learning offers a practical mechanism to counter these limitations by aggregating predictions from multiple independently adapted models. The core idea is that heterogeneous base learners capture different error patterns. When combined, their complementary strengths can reduce variance and mitigate systematic, per-class biases, thereby improving overall robustness~\cite{dietterich2000ensemble,zhou2025ensemble}. However, classical ensemble architectures can impose substantial computational burdens: training and serving many large-scale LLMs becomes infeasible under limited compute, making efficiency a key consideration in our ensemble design~\cite{hu2021lora}.


\begin{figure*}[htbp]
    \centering
    \includegraphics[width=1.1\linewidth]{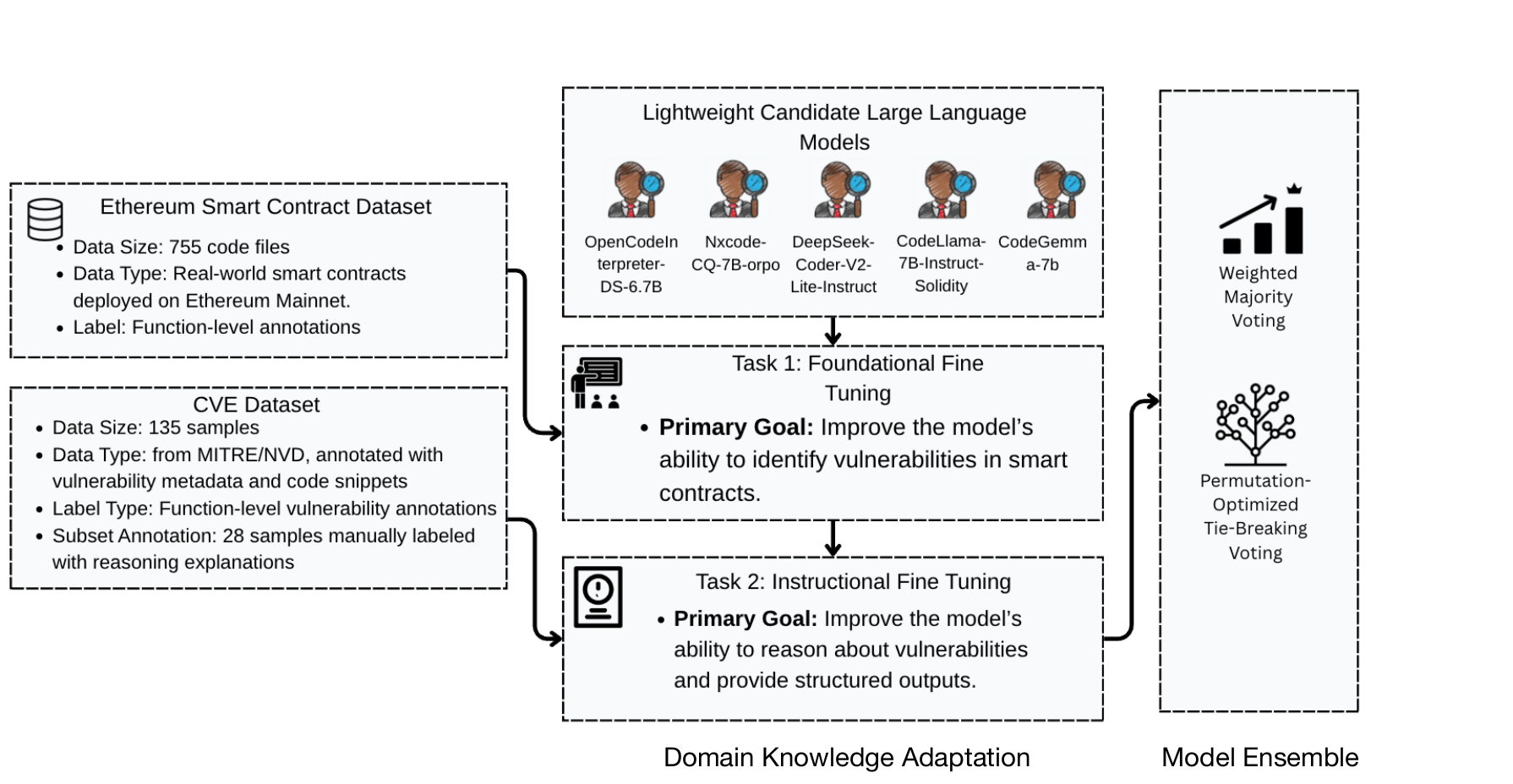}
    \caption{Overview of LLMBugScanner, which consists of two stages: domain knowledge adaptation and model ensemble to improve the effectivness of smart contract vulnerability detection.}
    \label{fig:method_diagram}
\end{figure*}

\section{Methodology}

As presented in Figure~\ref{fig:method_diagram}, we introduce LLMBugScanner to address the motivation described earlier. LLMBugScanner comprises two synergistic components: (1) domain knowledge adaptation, (2) model ensemble.

\begin{figure}[htbp]
    \centering
    \begin{subfigure}[b]{0.48\linewidth}
        \centering
        \includegraphics[width=\linewidth]{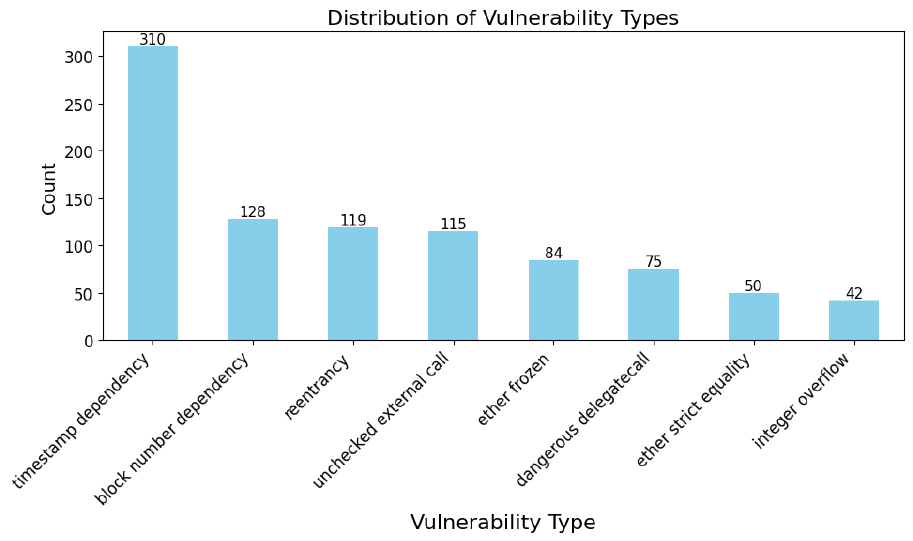}
        \caption{Distribution of vulnerability types in the Ethereum Smart Contract dataset for foundational fine-tuning. Timestamp dependency is the most prevalent.}
        \label{fig:eth_dataset}
    \end{subfigure}
    \hfill
    \begin{subfigure}[b]{0.48\linewidth}
        \centering
        \includegraphics[width=\linewidth]{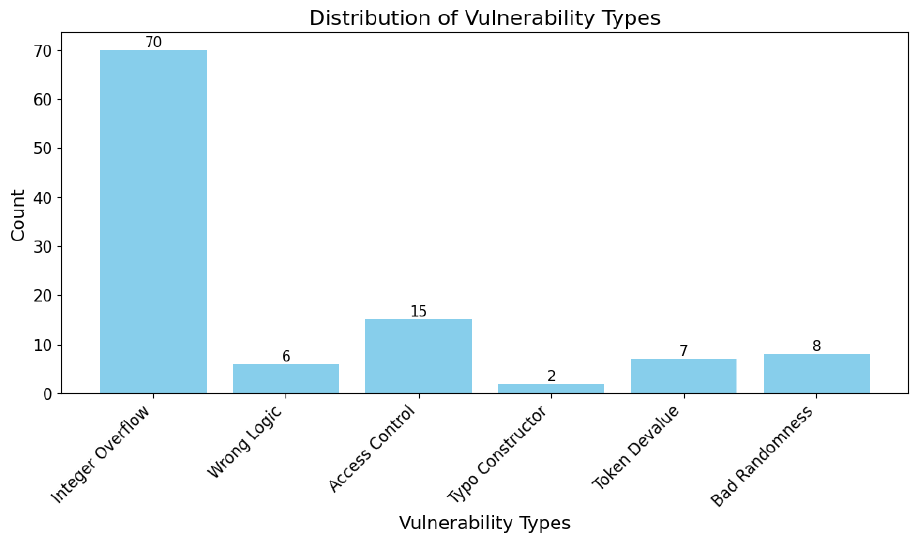}
        \caption{Distribution of vulnerability types in the CVE dataset used for instruction finetuning, with Integer Overflow being the most prevalent.}
        \label{fig:cve_dataset}
    \end{subfigure}
    \caption{Comparative distributions of vulnerability categories across the Ethereum and CVE datasets.}
    \label{fig:datasets_comparison}
\end{figure}

\textbf{Domain Knowledge Adaptation.} To broaden the coverage of vulnerabilities to reduce bias, we adopt two real-world datasets of smart contracts. The details of two datasets are shown in Figure~\ref{fig:datasets_comparison}:
\begin{itemize}
    \item \textbf{Ethereum Smart Contract Dataset:} 775 Solidity files labeled with vulnerabilities, used to improve general code interpretation. 
    \item \textbf{CVE Solidity Subset:} 28 files (25\% of full dataset), each annotated with a single vulnerability, used for instructional fine-tuning.
\end{itemize}

For each dataset, labels are transformed into structured prompts for the fine-tuning process. For the Ethereum Smart Contract dataset, contract names and vulnerability lists are extracted. For the CVE dataset, vulnerability type, function name, and description are included. Variables are dynamically replaced with actual dataset values during fine-tuning. To improve computational efficiency, we employ LoRA~\cite{hu2021lora}, which injects low-rank trainable adapters into attention and feed-forward projection layers, enabling efficient fine-tuning with only a fraction of parameters modified~\cite{hu2021lora}.

For fine-tuning, the auditor prompt specifies (1) task instructions with a restricted set of vulnerability types, (2) the number of vulnerabilities to return, (3) a JSON output schema with error handling, and (4) the full smart contract code as input, shown as in Figure~\ref{fig:auditor_prompt}.

\begin{figure}[tbp]
\small
\begin{tikzpicture}
  \node[draw, fill=gray!10, rectangle, rounded corners, inner sep=10pt] (box) {
    \begin{minipage}{1\textwidth}
      \textbf{Task Instructions:}\\[0.2em]
      Requirement: You are a smart contract auditor. Identify \{topk\} most severe vulnerability in the provided smart contract. Ensure it is exploitable in real world and beneficial to attackers. Restrict your identification to these vulnerability types: Integer Overflow, Wrong Logic, Bad Randomness, Access Control, Typo Constructor, Token Devalue.\\[0.5em]
      \textbf{Output Format:}\\[0.2em]
      You should ONLY output in below JSON format:
\begin{verbatim}
{
  "output_list": [
    {
      "function_name": "<func name>",
      "vulnerability": "<short desc>",
      "reason": "<reason>"
    },
    {
      "function_name": "<func name>",
      "vulnerability": "<short desc>",
      "reason": "<reason>"
    }
  ]
}
\end{verbatim}
      \textbf{Note:} If no vulnerabilities are found, output an empty JSON:
\begin{verbatim}
{"output_list": []}
\end{verbatim}
      \textbf{Full Code Input:}\\
      \{code\}\\[0.3em]
      \textbf{Your Output:}
    \end{minipage}
  };
\end{tikzpicture}
\vspace{-0.6cm}
\caption{Auditor prompt}
\vspace{-0.3cm}
\label{fig:auditor_prompt}
\end{figure}

\textbf{Model Ensemble.} We select top-performing models from the OpenCodeLLM leaderboard for parameter-efficient fine-tuning:
\begin{itemize}
    \item AlfredPros/CodeLlama-7b-Instruct-Solidity~\cite{AlfredPros_CodeLLaMa7BInstructSolidity_2024}
    \item m-a-p/OpenCodeInterpreter-DS-6.7B ~\cite{Zheng_IntegratingCodeGenerationWithExecution_2024}
    \item NTQAI/Nxcode-CQ-7B-orpo ~\cite{NTQAI_NxcodeCQ7Borpo_2025}
    \item deepseek-ai/DeepSeek-Coder-V2-Lite-Instruct (15.7B)~\cite{DeepSeek_CoderV2_BreakingBarrier_2024}
    \item google/codegemma-7b~\cite{Google_CodeGemma_2024}
\end{itemize}

Individual LLM predictions are aggregated through multiple ensemble techniques. These strategies allow us to exploit complementary strengths of different models and to reduce noise from individual misclassifications.

\paragraph{Method 1: Weighted Majority Voting.} We assign each model a weight based on its rank or performance (e.g., higher-performing models get higher weights). For each predicted (vulnerability, function) pair, the model's vote is multiplied by its weight. These weighted votes are then summed across all models. The final score for a pair \(j\) is:
\[
\text{Score}_j = \sum_{i \in M} w_i \cdot v_{ij}
\]
where \(M\) is the set of models, \(w_i\) is the weight assigned to model \(i\), and \(v_{ij} \in \{0,1\}\) indicates whether model \(i\) voted for pair \(j\). We then select the top five pairs with the highest total weighted scores as the ensemble predictions. This approach emphasizes high-quality models while still benefiting from consensus.

\paragraph{Method 2: Permutation-Optimized Tie-Breaking Voting.} 
We also evaluate a permutation-optimized, tie-breaking voting strategy. Here, we first compute unweighted votes:
\[
\text{Score}_j = \sum_{i \in M} v_{ij}
\]
If multiple pairs tie in \(\text{Score}_j\), the tie is broken using a learned model priority order \(\pi(i)\) (lower is higher priority). This order is optimized on a validation set to maximize ensemble accuracy on a held-out subset. Ties in \(\text{Score}_j\) are broken by the earliest model (lowest \(\pi(i)\)) that voted for the pair. This mechanism preserves the simplicity of unweighted voting while leveraging model-specific knowledge to resolve ambiguous cases.

In practice, we apply both methods and compare results on a validation set to choose the optimal ensemble for final deployment. Weighted voting tends to improve precision by giving more influence to strong models, while permutation-optimized tie-breaking improves recall by resolving conflicts in favor of historically reliable models. We also experiment with different values of \(k\) (number of top pairs retained) to assess trade-offs between coverage and accuracy.


\begin{figure*}
    \centering
    \includegraphics[width=1\linewidth]{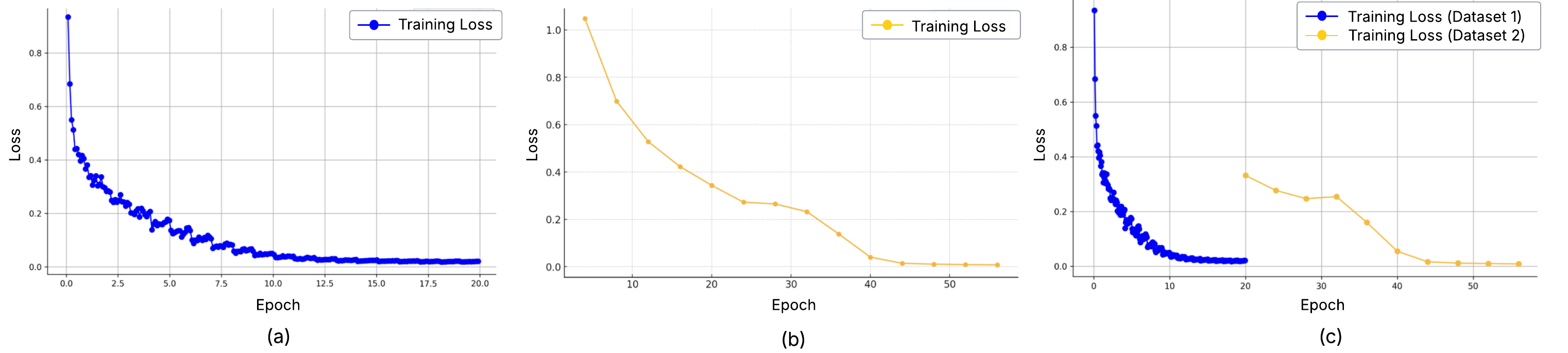}
    \caption{Training loss curves for models fine-tuned on Ethereum only (a), CVE only (b), and both datasets (c). The plots show rapid loss reduction during the early epochs with continued convergence across all settings. Notably, when fine-tuned on both datasets, the loss for the second dataset starts lower than in the standalone training scenario, highlighting the effectiveness of pre-training on the first dataset.}
    \label{fig:ft_effect}
\end{figure*}

\section{Experiment}
\label{sec:exp}

In this section, we validate the previous analysis and the efficacy of  LLMBugScanner via experimental results.

\subsection{Experimental Settings}

\textbf{Dataset:} We evaluate LLMBugScanner on a 80\% random split of CVE-Solidity dataset~\cite{hu2023large}, which contains 108 annotated smart contract files with each labeled with a single vulnerability type. The dataset captures diverse Ethereum smart contract vulnerabilities and serves as the benchmark for all model comparisons, shown in Fig.~\ref{fig:cve_dataset}.

\textbf{Models and Fine-Tuning:}  
Five open-source  light-weight large language models (LLMs) were selected based on their performance on code understanding tasks:  
AlfredPros/CodeLlama-7b-Instruct-Solidity, m-a-p/OpenCodeInterpreter-DS-6.7B, NTQAI/Nxcode-CQ-7B-orpo, deepseek-ai/DeepSeek-Coder-V2-Lite-Instruct, and google/codegemma-7b.  
Each model was fine-tuned using LoRA~\cite{hu2021lora} on two complementary datasets: the Ethereum smart contract dataset (general code interpretation) and a 25\% subset of the CVE dataset (instructional fine-tuning). LoRA hyperparameters were chosen based on reported best-performing settings in prior work. All models were evaluated in bfloat16 precision.

As shown in Fig.~\ref{fig:ft_effect}, models first fine-tuned on the Ethereum dataset exhibited substantially lower initial loss and faster convergence when subsequently trained on the CVE subset, compared to models trained on either dataset independently. This indicates that knowledge gained from the domain-general code representation phase transfers effectively to security-specific instruction tuning, reinforcing the benefit of sequential fine-tuning for cross-datasetss generalization.


\paragraph{Fine-Tuning Configuration.} 
All models are fine-tuned sequentially (Ethereum Smart Contract dataset → CVE subset dataset) using LoRA. The optimal hyperparameters were determined through a series of experiments varying learning rates, LoRA ranks, and batch sizes, with the final configuration selected based on validation performance and training stability.

\begin{table}[h]
\caption{LoRA Fine-Tuning Configuration (NXcode as Example)}
\centering
\begin{tabular}{ll}
\hline
\textbf{Component} & \textbf{Value} \\
\hline
Base Model & NTQAI/Nxcode-CQ-7B-orpo \\
LoRA Rank ($r$) & 32 \\
LoRA Modules & Attention + MLP + LM Head \\
Precision & bfloat16 \\
Batch Size & 32 \\
Epochs & 40 \\
Learning Rate & 2e-4 \\
Gradient Accumulation & 2 \\
Trainer & \texttt{trl.SFTTrainer} \\
\hline
\end{tabular}

\end{table}

\textbf{Auditor Prompt:}  
For evaluation, we employ a code auditing prompt that specifies (1) the role of the model as a vulnerability auditor, (2) the exact number of vulnerabilities to return, and (3) a strict JSON output format with error handling. The full smart contract code is provided as input in Fig.~\ref{fig:auditor_prompt}.

\textbf{Experiment Settings:}  
To compare auditing strategies, we implemented multiple configurations:  
\begin{itemize}
    \item \textbf{Single-Model Baselines:} Each LLM independently identifies vulnerabilities using the auditor prompt.
    \item \textbf{Single-Model Finetuned:} Each fine-tuned LLM independently identifies vulnerabilities using the auditor prompt.
    \item \textbf{Ensemble-Model Weighted Voting:} Models are assigned weights based on their individual top-$k$ hit rates, and the highest-scoring vulnerability predictions are selected.
    \item \textbf{Ensemble-Model Permutation-optimized Tie-Breaking Voting:} Permutation-optimized model ordering is used to break ties after unweighted voting, improving ensemble performance.
\end{itemize}

\textbf{Evaluation Metrics:}  
We report Top-$k$ hit rates under a strict match criterion (function name and vulnerability type) and under a soft cosine-similarity match criterion between model-generated descriptions and ground-truth labels at thresholds 0.5, 0.7, and 0.9. Both Top-1 and Top-5 hit rates are presented.

\textbf{Implementation Details:} All experiments were run on NVIDIA H100 GPUs. 
For all models, the generation parameters are set as shown in Table~\ref{tab:params}.

\begin{table}[t!]
\centering
\small
\caption{Parameter settings for generation}
\begin{tabular}{l l}
\toprule
\textbf{Parameter} & \textbf{Value} \\
\midrule
max\_new\_tokens & 800 \\
temperature & 0.1 \\
top\_k & 10 \\
top\_p & 0.95 \\
num\_return\_sequences & 1 \\
repetition\_penalty & 1.5 \\
\bottomrule
\end{tabular}
\label{tab:params}
\end{table}

This setup enables direct comparison between single-model performance and ensemble strategies, quantifying the gains from fine-tuning and ensembling on smart contract vulnerability detection.



\begin{table}[tbp]
\centering
\scriptsize
\setlength{\tabcolsep}{1pt}
\renewcommand{\arraystretch}{1}
\caption{Model Performance Comparison. For rows with cosine similarity (\textit{cs}), a hit is counted when the description appears in the Top-1 or Top-5 and its similarity exceeds threshold $t \in \{0.5, 0.7, 0.9\}$. 
The last two rows show direct-match hit rates (no threshold). 
FT = finetuned. 
Overall, finetuned models (FT) consistently outperform their baselines across most thresholds, indicating improved retrieval precision after finetuning. 
Notably, ensemble methods achieve the highest Top-5 hit rates, suggesting complementary strengths among individual models.
}

\label{tab:baseline-vs-finetuned}
\resizebox{\textwidth}{!}{
\begin{tabular}{l c *{16}{c}}
\toprule
\textbf{Metric} & \textbf{\(t\)} 
& \multicolumn{2}{c}{\textbf{Code Llama}} 
& \multicolumn{2}{c}{\textbf{DeepSeek}} 
& \multicolumn{2}{c}{\textbf{Gemma}} 
& \multicolumn{2}{c}{\textbf{NXCode}} 
& \multicolumn{2}{c}{\textbf{OpenInterpreter}} 
& \multicolumn{2}{c}{\textbf{Perm-Opt Ensemble}} 
& \multicolumn{2}{c}{\textbf{Weighted Ensemble}} \\
\cmidrule(lr){3-4} \cmidrule(lr){5-6} \cmidrule(lr){7-8} \cmidrule(lr){9-10} \cmidrule(lr){11-12} \cmidrule(lr){13-14} \cmidrule(lr){15-16}
 & & \textbf{Base} & \textbf{FT} 
   & \textbf{Base} & \textbf{FT} 
   & \textbf{Base} & \textbf{FT} 
   & \textbf{Base} & \textbf{FT} 
   & \textbf{Base} & \textbf{FT} 
   & \textbf{Base} & \textbf{FT} 
   & \textbf{Base} & \textbf{FT} \\
\midrule
Top 1 hit (cs) & 0.5 & 0.20 & 0.41 & 0.38 & 0.56 & 0.15 & 0.37 & 0.26 & 0.47 & 0.20 & 0.59 & 0.28 & 0.33 & 0.28 & 0.25 \\
Top 5 hit (cs) & 0.5 & 0.44 & 0.43 & 0.77 & 0.82 & 0.37 & 0.50 & 0.68 & 0.47 & 0.50 & 0.73 & 0.69 & 0.73 & 0.69 & 0.84 \\
Top 1 hit (cs) & 0.7 & 0.05 & 0.29 & 0.12 & 0.22 & 0.02 & 0.29 & 0.09 & 0.34 & 0.04 & 0.45 & 0.06 & 0.16 & 0.05 & 0.07 \\
Top 5 hit (cs) & 0.7 & 0.06 & 0.32 & 0.20 & 0.33 & 0.05 & 0.37 & 0.31 & 0.34 & 0.18 & 0.49 & 0.32 & 0.49 & 0.31 & 0.34 \\
Top 1 hit (cs) & 0.9 & 0.00 & 0.08 & 0.00 & 0.00 & 0.00 & 0.21 & 0.00 & 0.06 & 0.00 & 0.24 & 0.00 & 0.05 & 0.00 & 0.00 \\
Top 5 hit (cs) & 0.9 & 0.00 & 0.09 & 0.00 & 0.00 & 0.00 & 0.21 & 0.00 & 0.06 & 0.00 & 0.25 & 0.00 & 0.20 & 0.00 & 0.00 \\
Top 1 hit (direct) & – & 0.06 & 0.27 & 0.05 & 0.38 & 0.04 & 0.28 & 0.09 & 0.34 & 0.06 & 0.39 & 0.07 & 0.40 & 0.06 & 0.093 \\
Top 5 hit (direct) & – & 0.12 & 0.31 & 0.14 & 0.56 & 0.11 & 0.38 & 0.41 & 0.34 & 0.27 & 0.41 & 0.44 & 0.60 & 0.44 & 0.56 \\
\bottomrule
\end{tabular}
}
\end{table}

\subsection{Performance Comparison}
The results are presented in Table~\ref{tab:baseline-vs-finetuned}, where we can make several observations:
\begin{itemize}
    \item \textbf{Fine-tuning boosts Top-hit performance.} Across almost all models and thresholds, fine-tuned versions achieve substantially higher Top-1 and Top-5 hit rates than their baselines. Gains are most pronounced at the moderate similarity threshold \(t=0.5\), showing that the fine-tuned models produce more relevant matches under relaxed and moderate criteria.
    
    \item \textbf{Permutation-optimized vs. weighted ensemble.} The permutation-optimized ensemble attains consistently higher Top-1 rates than individual models at \(t=0.5\)–\(0.7\), whereas the weighted ensemble shows a clear advantage in Top-5 metrics (e.g., 0.84 at \(t=0.5\) vs.\ 0.73 for perm-opt). This suggests the perm-opt method prioritises precision on the single best candidate, while the weighted scheme increases recall across several candidates, boosting Top-5 performance.
    
    \item \textbf{Description similarity thresholds magnify differences.} As the cosine-similarity threshold increases from 0.5 to 0.9, hit rates collapse for all models—especially for Top-1. However, fine-tuned models and both ensemble variants retain a relative advantage even at \(t=0.7\)–\(0.9\), implying their descriptions are not only more likely to appear in the ranked list but also more semantically aligned with the ground truth.
    
    \item \textbf{Direct match versus similarity-based match comparison.} The ensemble model steadily improves direct match (by function and vulnerability name) in both top-1 and top-5 accuracy. However, in the description similarity match, it sometimes performs worse than the top-performing single model, in this case, \textit{open-interpreter}. This indicates that there may be a misalignment between the models’ textual reasoning and their actual vulnerability/function identification, suggesting that some ensemble mechanisms amplify shallow consensus rather than deeper semantic understanding. Moreover, The ensemble could be overfitting to lexical overlap in predictions, leading to improved direct matching but weaker semantic alignment. Last, Incorporating semantic-aware aggregation (e.g., embedding-based or explanation-consistency weighting) might help align reasoning quality with identification accuracy in future iterations.
    
\end{itemize}

To summarize, these results suggest that fine-tuning and ensembling provide complementary advantages. Fine-tuning improves each model in detecting frequently occurring vulnerabilities and produces more semantically consistent outputs, while the ensemble, especially the weighted version, broadens coverage across multiple plausible predictions, which is valuable for retrieval tasks where several closely matching answers can all be acceptable.

\subsection{Case Study: best single model vs. best ensembled model} 
To move beyond aggregate metrics, we performed a case analysis comparing the best-performing single model with the ensemble model across four correctness scenarios: (1) both models incorrect, (2) best model correct but ensemble incorrect, (3) ensemble correct but best model incorrect, and (4) both models correct. This analysis does not simply describe the distributions of vulnerability types and functions, but reveals the mechanisms by which the ensemble improves or degrades detection performance.

\begin{figure}[!htbp]
    \centering
    \vspace{-5mm}
    \includegraphics[width=0.8\linewidth]{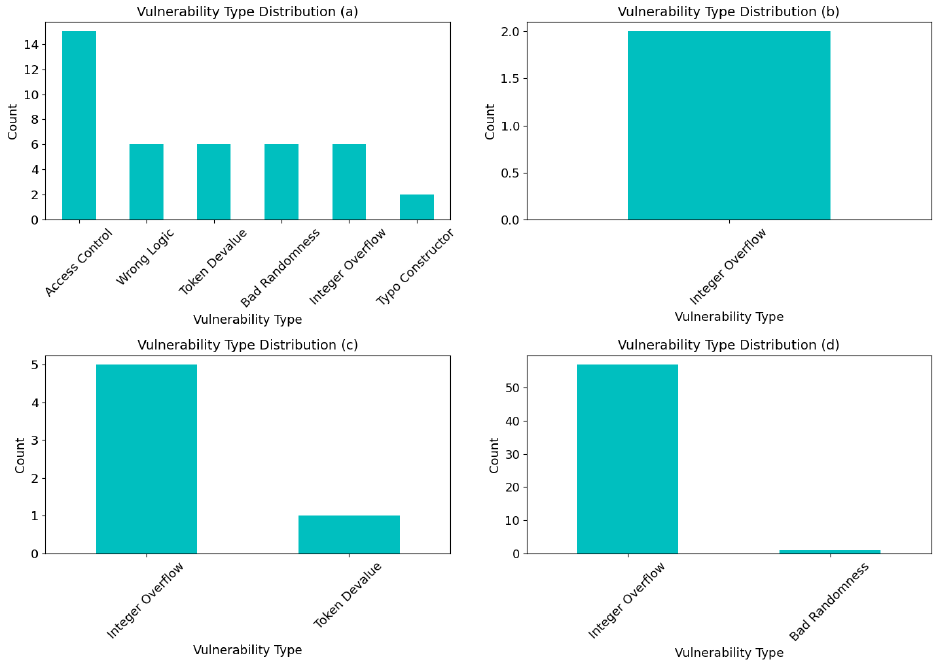}
    \caption{Vulnerability-type distributions under different model agreement scenarios. (a) Cases where both the ensemble and DeepSeek are incorrect show a wider spread across vulnerability types, with Access Control dominating the errors. (b) Cases where DeepSeek is correct but the ensemble is incorrect are rare and concentrated on a single vulnerability type (Integer Overflow). (c) Cases where DeepSeek is incorrect but the ensemble is correct indicate that ensemble predictions recover some errors from DeepSeek, mainly on Integer Overflow and Token Devalue vulnerabilities. (d) Cases where both the ensemble and DeepSeek are correct are dominated by Integer Overflow, with a small fraction of Bad Randomness vulnerabilities, showing the strongest area of agreement.}
    \vspace{-10mm}
    \label{fig:vuln-type}
\end{figure}

\paragraph{Recovering High-Impact Vulnerabilities.}
A key insight from Figure~\ref{fig:vuln-type} is that the ensemble recovers some vulnerabilities, particularly Integer Overflow and Token Devalue, that the best single model misses. However, there are two Integer Overflow cases missed where the best single model correct. The fact that Integer Overflow is the dominant case could lead to overfitting for all the models. This could explain the increase in correctly identified Integer Overflow cases by the ensemble model. When combined with the two Integer Overflow cases miss and two Token Devalue cases hit for ensemble model, ensemble model shows an ability to provide a less biased prediction while could lead to some mismatch and introduce some noise. On the other hand, both models have a hard time predicting other minority classes include Wrong Logic, Access Control, and Typo Construction. That shows that the ensemble model still cannot handle minority cases when the training set is limited and biased due to each model in the ensemble model faces same problem. One important vulnerabilities is Access Control. With a relative large sample number compared to Token Devalue and Bad Randomness, the models had a hard time to make a correct prediction in this specific type of vulnerabilities. Since each model are pre-trained with larger dataset, the problem could be the knowledge from the pre-training dataset are not enough for model to make correct decision. It could also imply that the Access Control cases are more complicated, and hard to be caught by current model we have.

\begin{figure}[!tbp]
    \centering
    \vspace{-3mm}
    \includegraphics[width=0.8\linewidth]{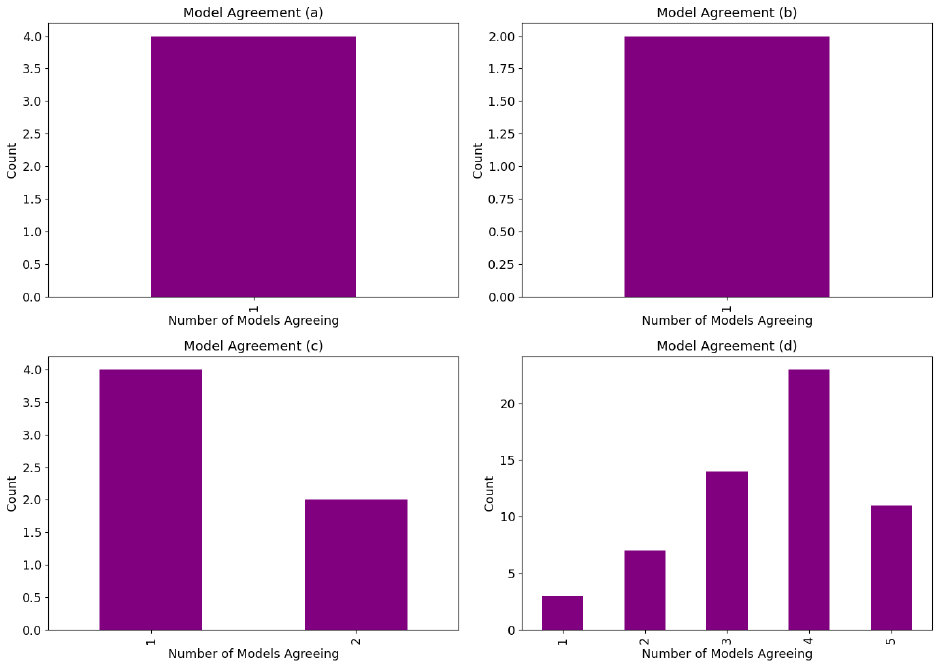}
    \caption{Model agreement distribution under different model agreement scenarios. (a) Both the ensemble and DeepSeek are incorrect. (b) DeepSeek is correct but the ensemble is incorrect. (c) DeepSeek is incorrect but the ensemble is correct. (d) Both ensemble and DeepSeek are correct, showing consistently high agreement (4–5 models), demonstrating that strong consensus within the ensemble correlates with accurate predictions.}
    \vspace{-5mm}
    \label{fig:voting}
\end{figure}

\paragraph{Voting Distribution and Consensus Dynamics.}
The model agreement plots (Figure~\ref{fig:voting}) quantify voting dynamics. When both models are wrong, agreement within the ensemble is minimal, indicating that diversity alone does not guarantee success on truly hard cases. Especially for minority case, the nature of lack of training data make all models have a hard time to make good prediction. In this case, the ensemble itself cannot improve the performance of model prediction. When the best model is correct but the ensemble is wrong, agreement is also low, reflecting the dilution of a correct minority vote. By contrast, when the ensemble is correct and the single model is wrong, moderate agreement emerges; here, the ensemble is successfully amplifying weak but consistent votes across models. When both models are correct, consensus is highest, reflecting that easy or obvious cases drive strong agreement.

Overall, these observations provide two actionable insights for improving ensemble-based vulnerability detection. First, ensembles deliver the largest gains on high-severity, heterogeneous vulnerability types and functions, suggesting that model diversity is most valuable where classification requires reasoning across multiple semantic cues. Second, ensembles are vulnerable to overcorrection in low-diversity settings, implying that weighting or confidence calibration of constituent models could further improve performance.


\section{Future Directions}

\paragraph{Learning-Based Ensembles.}
While our current architecture uses rule-based ensembles over multiple fine-tuned LLMs, a more advanced path is to adopt a learning-based ensemble strategy. Instead of fixed rules to decide which model’s output to trust, one could train a \textit{meta-classifier} that, given features of the input, predicts which LLM is most likely to produce a correct vulnerability diagnosis. This approach essentially turns the ensemble step into a learned decision problem. To do this reliably, one would require a larger labelled dataset: the data would have to be partitioned not just into train/validation/test, but also into a classifier-training split (for learning which model to trust) and a held-out final test split to evaluate end-to-end performance. The overhead is nontrivial, but this approach has been successful in other ensemble settings (e.g., stacking in machine learning)~\cite{ridoy2024enstack,wolpert1992stacked} and could adapt dynamically to different types of smart contract code.

\paragraph{Hallucination Mitigation.}
A second direction is to tackle hallucination and inconsistent outputs. In our experiments, we observe a $\sim$10\% hallucination rate across models—instances where the model invents vulnerabilities or outputs contradictory diagnoses. This likely stems from using lightweight fine-tuning with constrained adaptation capacity. Future work could mitigate hallucinations via several strategies: (i) \textit{contrastive regularization}, where known safe code is paired explicitly against vulnerable code to penalize stray predictions~\cite{ji-etal-2024-applying}; (ii) \textit{self-consistency verification}, where multiple prompting or sampling is used and only consensus outputs are accepted~\cite{wang2022selfconsistency,chen2023universal}; and (iii) integrating \textit{symbolic consistency checks} as a filter layer—e.g., verifying that a flagged vulnerability is consistent with the control-flow or state transitions of the contract code. Combining LLM predictions with symbolic validators may reduce hallucination while retaining flexibility. Further, self-reflective frameworks such as SaySelf~\cite{xu2024sayself} could be explored to help models estimate their own confidence and suppress low-certainty predictions.

\paragraph{Code Normalization and Canonicalization.}
A third promising direction is to enhance model robustness through code normalization and canonicalization. Variation in variable names, formatting, ordering of statements, or syntactic sugar can make it harder for an LLM to generalize vulnerabilities across code that is semantically equivalent but syntactically different. Introducing a normalization pre-processing step—such as renaming variables to canonical tokens, rewriting expressions to a normalized intermediate form, or applying AST reordering—could reduce the noise the model sees. Prior work on code normalization and embedding learning has shown improved generalization for downstream tasks~\cite{kanade2020learning,zhang2023codet5plus,ahmad2023code}. One could even train a denoising autoencoder (or fine-tuned model) that maps arbitrary Solidity code into a canonical form before vulnerability detection. Doing so may boost the models’ robustness to superficial syntactic variation and improve generalization to unseen contract styles.

In summary, advancing our system along these paths—meta-learned ensembles, hallucination filtering, and normalization pipelines—offers a roadmap to more accurate, credible, and robust smart contract vulnerability detection.

\section{Conclusion}


We have presented LLMBugScanner, a systematic LLM-powered approach for smart contract vulnerability detection. The design of LLMBugScanner has several unique features. 
First, we effectively utilize domain knowledge adaptation through multi-stage fine-tuning across different benchmark datasets with parameter-efficient techniques. Second, we integrate domain-knowledge adaption with LLM-ensemble to enable LLMBugScanner to acquire complementary expertise across multiple independently pre-trained and fine-tuned LLMs over multiple general programming and security-specific datasets. Our LLM-ensemble method resolves reasoning conflicts among multiple LLMs through consensus convergence. 
Experimental results across multiple LLM families, including CodeLlama, DeepSeek, NXCode, and CodeGemma, show that LLMBugScanner offers consistent improvements in precision, recall, and robustness over single-model baselines studied in this paper, highlighting LLMBugScanner as a cost-effective, extensible, and trustworthy approach for LLM-based smart contract auditing.

\section*{Acknowledgements}

This research is partially sponsored by the NSF CISE grants 2302720 and 2312758, an IBM faculty award, and a CISCO research grant in Edge AI.

\bibliographystyle{splncs04}  
\bibliography{reference}     

\end{document}